\documentstyle[psfig]{kapproc} 

\kluwerbib

\startauthorindex

\begin{document}

\articletitle{Non coeval young multiple systems?}

\articlesubtitle{On the pairing of protostars and T\,Tauri stars}

\author{Gaspard Duch\^ene, Andrea Ghez \& Caer McCabe}
\affil{UCLA -- Department of Physics \& Astronomy\\
Los Angeles, CA 90095-1562, USA}
\email{duchene@astro.ucla.edu, ghez@astro.ucla.edu,
mccabe@astro.ucla.edu}

\chaptitlerunninghead{Non coeval young multiple systems?}

\anxx{Author1\, and Author2}

\begin{abstract}
  We summarize here the observed properties of ``infrared companions''
  to T\,Tauri stars and argue that their observational properties are
  identical to those of Class\,I sources. They may therefore be
  embedded protostars in a much earlier evolutionary phase than
  T\,Tauri stars, in which case these multiple systems are
  significantly non-coeval as opposed to the majority of young binary
  systems. They would have formed through a different mechanism than
  core fragmentation. The only distinction between IRCs and Class\,I
  sources is that they lie within a few tens of AU of a T\,Tauri star,
  and so they cannot be at the center of a vast optically thick
  envelope as is believed to be the case for protostars. We discuss
  whether systems with an IRC are really candidates for non-coeval
  multiple systems.
\end{abstract}

\section{Introduction}

Binary and higher order multiple systems are the most frequent outcome
of the star formation process and, as such, they represent a direct
probe of this process. In the most widely accepted model to date,
giant molecular clouds first give rise to a number of individual
clumps, or cores, that subsequently undergo a collapse to finally form
a central star. During this collapse, fragmentation may occur, due to
rotation, turbulence or ambipolar diffusion of the magnetic field for
instance, that leads to the formation of $\sim$2--5 physically
associated objects which eventually evolve into a stable multiple
system. One of the basic predictions of this scenario is that binary
systems should be tightly coeval, to within a free-fall time of the
original core ($\ll 10^6$\,yrs). Observational campaigns to test this
prediction on large populations of T\,Tauri binary systems have been
conducted for several years (e.g., Hartigan et al. 1994; White \& Ghez
2001) and have concluded that binary systems are remarkably coeval, at
least to within $<10^6$\,years. Therefore, observations tend to favor
core fragmentation as the dominant process of forming binary stars.

Among all T\,Tauri binary systems, there is a small category of
remarkable objects named ``infrared companions'' (IRCs). These are
defined as companions to known T\,Tauri stars that can only be
detected in the infrared (IR) and that display ``extremely'' red IR
colors, probably because of a large line-of-sight extinction. The most
prominent IRC is the companion to the prototypical object T\,Tau (Dyck
et al. 1982). The status of this class of objects was first discussed
by Zinnecker \& Wilking (1992) and a good compilation of observations
can be found in Koresko et al. (1997), after which a few more IRCs
were identified. As we will discuss here, these systems are
problematic, as they seem to pair a normal T\,Tauri star with a much
more embedded, and qualitatively much younger, protostar. They would
therefore be non-coeval systems, contrasting with most young binary
systems.

Since the review by Koresko et al., additional high spatial resolution
data have been obtained and information about the dynamics of the
systems have been gathered by us and others.  Here, we summarize those
results and discuss how they fit into the binary coevality issue.

\section{Observational properties of IRCs}

\subsection{A naive look at IRCs: an age paradox}

In the well-studied Taurus-Auriga star-forming region, 5 IRCs are
known among a sample of $\sim120$ stars, representing a mere 4\,\% of
the overall population. The masses and ages of their optically bright
primaries are in the ranges $\sim$0.2--2\,$M_\odot$ and
$\sim$0.1--5\,Myrs respectively.  The projected separations range from
$\sim10$ to a few hundred AUs, similar to the range of separations for
normal T\,Tauri binaries.  Also, most systems show thermal millimeter
emission associated with cold circumstellar material although in
general it is not known with which component it is associated.  The
fact that some of these IRCs have (at times) been detected in the
visible raises the possibility that intrinsically different objects
are erronneously assembled in the same category.

As a first step to determining the nature of IRCs, Koresko et al.
(1997) compiled the spectral energy distribution (SED) for each IRC
from the visible up to 100\,$\mu$m. They all showed a similar shape,
with a very broad peak centered between 5 and 20\,$\mu$m depending on
the object. The corresponding bolometric temperatures are much cooler
than any stellar photosphere, only a few hundred degrees.  Assuming
that they radiate isotropically, the bolometric luminosities of IRCs
range from $\sim$0.8 to 12\,$L_\odot$. Taking these values at face
value and plotting them against other young stellar objects as well as
against the protostar evolution models of Myers et al. (1998), one
finds that IRCs fall in the same part of the diagram as Class\,I
sources, protostars embedded in a moderately massive, contracting
circumstellar envelope. Therefore, if one were to classify IRCs on the
basis of their SED only, one would undoubtedly conclude that IRCs are
Class\,I protostars.

T\,Tauri stars in general, and those that have an IRC in particular,
have typical ages of 1--5\,Myr. On the other hand, it is generally
admitted that the small numbers of Class\,I protostars implies that
they are much younger, typically a few $10^5$\,yrs. It therefore
appears that IRC systems are non-coeval, with multi-Myr age
differences. Three general explanations can be proposed to account for
these systems: i) some multiple systems are really non-coeval, in
which case one must explain how they formed; ii) IRCs only {\it look
  like} protostars but are in fact T\,Tauri stars disguised as
protostars; and iii) the T\,Tauri stars associated to IRCs are in fact
much younger than we think they are and these systems are in fact
extremely young, coeval systems.

\subsection{A purely geometrical explanation?}

Among the three explanations suggested above, the idea that IRCs are
in fact normal T\,Tauri stars in a peculiar geometric configuration is
the most widely believed (e.g., Koresko et al. 1997). Such objects
would look like protostars if they were heavily extincted by some
circumstellar material, as none or very little flux shortward of
1\,$\mu$m would reach the observer while at long wavelengths one would
detect the thermal emission of the dusty material that enshrouds the
central star.

There are two types of configurations that would lead to the observed
SEDs for IRCs. The first one is the case of a star that is embedded in
an optically thick dusty envelope so that the only light we receive
from the star has been reprocessed. Alternatively, IRCs could be
T\,Tauri stars surrounded by an unresolved optically thick
circumstellar disk that is seen at an almost edge-on inclination.
Both observations and radiative transfer models of such objects show
that their SED is extremely suppressed shortward of $\sim10\,\mu$m,
resulting in a predominant mid-IR peak, similar to those of Class\,I
sources (D'Alessio et al. 1999; Wood et al. 2002). In this case, the
inferred bolometric luminosity assuming isotropic radiation can be
1--2 orders of magnitudes lower than its actual value because the
visible/near-IR light is predominantly scattered away from the
observer's line of sight.

A relatively straightforward observational test can discriminate the
two scenarios presented here: if the star is embedded into an
optically thick envelope, its near-IR (and visible if observed)
spectrum is featureless as it has been reprocessed by warm dust. On
the other hand, in the edge-on disk scenario, the received spectrum
has only been scattered at the surface of the disk and has retained
the intrinsic photospheric features of the central object. This has
for example been verified for the edge-on disk source IRAS\,04158+2805
(M\'enard et al. 2003).

\section{Recent observations of IRCs}

Over the last few years, at least two new IRCs have been identified,
WL\,20\,S and V\,773\,Tau\,D (Ressler \& Barsony 2001, Duch\^ene et
al. 2003), and several high angular resolution datasets have been
obtained, both in imaging and spectroscopic modes, for several
systems, allowing a more complete understanding of their properties.

First of all, IRCs are extremely variable, by up to several magnitudes
even in the mid-IR. This was already known for some of them (e.g.,
T\,Tau: Ghez et al. 1991) and has also been observed for
V\,773\,Tau\,D (Duch\^ene et al. 2003). Also strong absorption
features of both water ice and silicates have been observed in the IR
(Hanner et al. 1998, Beck et al. 2001). These features unambiguously
show that IRCs are observed through large column densities of dusty
material.  The photometric variability is however unlikely to be fully
explained by a varying line-of-sight extinction: variations in the
emission of the central source has to be present as well (Beck et al.
2001; Leinert et al. 2001).

Recent near-IR spectroscopy of several IRCs (Haro\,6-10\,N, T\,Tau\,S,
V\,773\,Tau\,D) have revealed featureless spectra, with the exception
of atomic and molecular hydrogen in emission (Herbst et al. 1995; Beck
et al.  2001; Duch\^ene et al. 2002, 2003). This excludes the
possibility of these IRCs being K- or M-type T\,Tauri stars extincted
by an edge-on circumstellar disk. Note that this result is not
inconsistent with IRCs being earlier spectral type (i.e., higher mass)
objects seen behind an edge-on disk. This is discussed in more details
in the following section.

The case of T\,Tau\,S is quite revealing since this IRC is located in
a triple system (Koresko 2000). Most importantly, Duch\^ene et al.
(2002) have shown that the tight companion to the IRC, which is
located only 10--12\,AU away, is a normal, though heavily extincted,
T\,Tauri star with an M0.5 spectral type. This implies that what makes
an IRC so special is confined within a few AU of the central object.
If it is an opaque circumstellar envelope, it has to be quite dense in
order to be optically thick despite such a small radius.

\begin{figure}[t]
\begin{tabular}{cl}
\psfig{figure=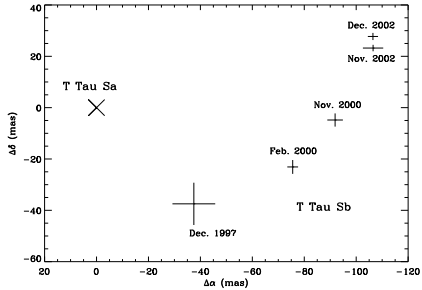,width=0.48\textwidth} &
\hspace*{0.1cm} 
\begin{minipage}[b]{0.48\textwidth}
\caption{\label{fig:ttaus} Location of the companion to the
  IRC of T\,Tau -- a.k.a. T\,Tau\,Sa -- as a function of time since it
  discovery. Data are from Koresko (2000), K\"ohler et al. (2000) and
  Duch\^ene et al. (2002). The two most recent points were obtained by
  us with the adaptive optics system on the Keck\,II telescope.}
\vspace*{0.5cm}

\end{minipage}
\end{tabular}
\vspace*{-0.5cm}

\end{figure}

Finally, the most exciting new result regarding IRCs concerns their
dynamical status. Most of them are located at a few hundred AU and the
orbital periods for those systems are on order of a thousand years.
However, T\,Tau\,S is in a 10\,AU-binary and we can expect its orbital
period to be about 15--30\,yrs. Since its first discovery in 1997,
several measurements of the binary separation have been made and clear
evidence of orbital motion has been found (see Fig.\,\ref{fig:ttaus}).
In November 2000, the relative velocity of the binary in the plane of
the sky was on order of 13$\pm$4\,km/s (based on a quadratic fit).
Furthermore, we have also obtained spatially resolved high spectral
resolution near-IR spectra of both components of the system.  We found
a relative velocity of 20$\pm$2\,km/s. This combines to a three
dimension relative velocity of about 24$\pm$4\,km/s which implies a
minimum system mass of $M_{TTau S}\geq(4.2\pm1.5)(\frac{D}{140\,{\rm
    pc}})^3 M_\odot$, if the system is physically bound. This is much
more than the estimated mass of T\,Tau\,Sb and T\,Tau\,N, suggesting
that the IRC is the most massive object in the T\,Tau multiple system.
So far, the measurements do not cover enough of the orbit to allow a
complete orbital solution fit but this should be feasible in just a
few years.

A controversial result concerning this system was recently obtained by
Loinard et al. (2003) using archival VLA centimeter data. At these
wavelengths, only T\,Tau\,N (the well known T\,Tauri star) and
T\,Tau\,Sb (the extincted close companion to the IRC) are detected.
Still, they concluded that the T\,Tau\,Sa--Sb system is unstable and
that that T\,Tau\,Sb has been ejected in the last few years on a
higher (possibly open) orbit. This is reminiscent of the ``disrupting
triple systems'' proposed by Reipurth (2000), which were candidates
for forming IRC systems. In this scenario, a very young (a few
$10^5$\,yrs-old) triple system undergoes an unstable triple encounter
and one of them is ejected in one direction while the remaining binary
experiences a slow recoil motion in the other direction. The single
star would then escape the opaque envelope still surrounding the
system and therefore become optically visible while the other two
components would remain heavily obscured (this is the third scenario
mentionned in \S\,2.1). However, this scenario is not very well
supported by our IR images, which show a clear slowdown of the motion
of T\,Tau\,Sb between 2000 and 2002, suggesting that its orbit is
bound. Only future observations will tell what type of orbit this star
is on.

In the case of V\,773\,Tau, the IRC has not been monitored long enough
to allow a proper orbital fit. However, all observations are
consistent with the system being hierarchical, as well as possibly
coplanar, and therefore dynamically stable. Most other IRCs are
located in binary systems and, unless a third component is later
discovered, they are necessarily in two-body stable systems. Until
further measurements prove otherwise, we conclude that IRCs are in
stable systems and that Reipurth's scenario is not generally the cause
for these unusual IRCs.

\section{The high-mass star hypothesis}

One of the scenarios presented above to account for the observed
properties of IRCs is that they are high- to intermediate-mass stars
obscured by a circumstellar disk seen edge-on. This is a likely
situation in Glass\,I, since its spectral type has been estimated to
be G5 (Feigelson \& Kriss 1989) but is clearly inconsistent with the
spectral type of UY\,Aur\,B (M2, Duch\^ene et al. 1999). In systems
with featureless IR spectra, the central star could be an A- or F-
type star, preventing any line detection if the star is accreting
material (accretion produces hydrogen line emission that fills
photospheric features).  Also, such objects would have large
luminosities, $\gg10\,L_\odot$, but only a fraction of it would be
seen by the observer because of the peculiar geometry.  This would
explain the observed luminosities of IRCs. Finally, in the case of
T\,Tau\,S, a relatively large mass is required for the IRC if the
system is physically bound. The high-mass star hypothesis is therefore
a significant possibility that needs to be studied in more details.

One way to test this hypothesis is to obtain high-resolution spectra
of IRCs and try to find some photospheric features in them. It is for
instance suggestive that, in our radial velocity measurement, we
obtained the strongest cross-correlation peak for T\,Tau\,Sa with the
spectrum of an F8 template star (we used M5 to F8 templates). A larger
set of templates is however required to determine accurately the
actual intrinsic spectral type of this IRC. Another possible test
consists in analyzing the spectrum of the scattered light nebula
surrounding T\,Tau. If the IRC is a 3--5\,$M_\odot$ star, it is by far
the most luminous object in the system and it should be the dominant
source of illumination for the nebula.

\section{Are IRCs bona fide protostars?}

If IRCs are not high-mass objects, then they have to be surrounded by
compact optically thick circumstellar envelopes. It is then natural
that they have similar properties as Class\,I protostars, as they are
virtually identical: a central point source surrounded by a lighter
but opaque envelope. In this scenario, one wonders why a T\,Tauri star
would be surrounded by an optically thick envelope, since they are
usually defined as objects around which the vast majority of the
circumstellar material lies in an equatorial disk. It has been
proposed that they are in fact deeply embedded because they are
experiencing a temporary high accretion rate event similar to FU\,Ori
bursts (Ghez et al.  1991, White \& Ghez 2001) so that their opaque
envelope would be a transient phenomenon. The number of IRCs would
then suggest a relatively short-lived phenomenon ($\approx
10^4$\,yrs). It remains to be understood how such an opaque envelope
suddenly appears around one of the components of the system. This may
be the result of star-star dynamics within the systems (Bonnell \&
Bastien 1992) but the details of this phenomenon still have to be
described.

There is however one usual property of protostars that IRCs do not
share: a vast envelope. Millimeter mapping has shown that protostars
have envelopes thousands of AU in radius (e.g., Motte \& Andr\'e 2001)
whereas the fact that IRCs are in multiple systems imply that the
outer radius of their envelope is not bigger than $\sim$2--100\,AU
depending on the system.  From this point of view, IRCs are unlikely
to be actual protostars. By analogy, this implies that Class\,I
objects in general, which are defined by their near- to mid-IR SED,
should not be considered protostars without further analysis, even
though they are not known to have a companion. In fact, as discussed
in Andr\'e et al. (2000), the presence of a massive and extended
envelope is required to consider an object a bona fide protostar. With
this criterion, none of the IRCs can be considered a real protostar
and the apparent non-coevality of the systems is solved.

In summary, this study reminds us, from an unusual perspective, that a
Class\,I-type SED is not the ultimate proof that an object is a
protostar and that there might be a few non-coeval multiple systems in
star-forming regions. If we take the Class\,I classification of IRCs
from their SED at face value, we then consider these systems as not
coeval and we must explain how they formed. One possibility is that
IRCs formed after a gravitational instability disrupted the
circumstellar disk of their companion. This could represent a
secondary channel for star formation and, although it only amounts to
a few percent of all the stars formed in an environment such as
Taurus, it would be interesting to see if different star-forming
regions can form more objects in this way. Alternatively, IRCs might
be FU\,Ori-like objects as old as their T\,Tauri companion, in which
case dynamical interactions within the systems would need to be
extremnely strong. Finally, a more original scenario for explaining
(some of the) IRCs is that they are high-mass objects extincted by a
nearly edge-on disk. If so, this would imply that even the Taurus
molecular cloud can form objects as massive $\approx5\,M_\odot$. In
any case, the unusual properties of IRCs deserve further
investigation.

\begin{acknowledgments}
  The authors are grateful to the organizers of the conference for its
  great atmosphere, highly interesting list of topics and professional
  organisation. Many thanks also to Hans Zinnecker, Nuria Calvet, Lee
  Hartmann, Bo Reipurth and Philippe Andr\'e for enriching discussions
  following this presentation. This work has been supported by the
  National Science Foundation Science and Technology Center for
  Adaptive Optics, managed by the University of California at Santa
  Cruz under Cooperative Agreement AST 98-76783.
\end{acknowledgments}


\begin{chapthebibliography}{1}
\bibitem{andre00}
Andr\'e, P., Ward-Thompson, D. \& Barsony, M. 2000, in
``Protostar \& Planets IV'', eds. Mannings, Boss \& Russell,
Univ. of Arizona Press, p. 59

\bibitem{beck01}
Beck, T., Prato, L. \& Simon, M. 2001, ApJ, 551, 1031

\bibitem{bonnell_bastien92}
Bonnell, I. \& Bastien, P. 1992, ApJ, 400, 579

\bibitem{dalessio99}
D'Alessio, P. {\it et al.} 1999, ApJ, 527, 893

\bibitem{duchene99}
Duch\^ene, G. {\it et al.} 1999, A\&A, 351, 954

\bibitem{duchene02}
Duch\^ene, G., Ghez, A. M. \& McCabe, C. 2002, ApJ, 568, 771

\bibitem{duchene03}
Duch\^ene, G. {\it et al.} 2003, ApJ, in press
(astro-ph/0303648)

\bibitem{dyck82}
Dyck, H. M., Simon, T. \& Zuckerman, B. 1982, ApJ, 255, L103

\bibitem{feigelson_kriss89}
Feigelson, E. D. \& Kriss, G. A. 1989, ApJ, 338, 262

\bibitem{ghez91}
Ghez, A. M. {\it et al.} 1991, AJ, 102, 2066

\bibitem{hanner98}
Hanner, M. S., Brooke, T. K. \& Tokunaga, A. T. 1998, ApJ,
502, 871

\bibitem{hartigan94}
Hartigan, P., Strom, K. M. \& Strom, S. E. 1994, ApJ, 427, 961

\bibitem{herbst95}
Herbst, T. M., Koresko, C. D. \& Leinert, C. 1995, ApJ, 444,
L93

\bibitem{koresko97}
Koresko, C. D., Herbst, T. M. \& Leinert, C. 1997, ApJ, 480,
741

\bibitem{koresko00}
Koresko, C. D. 2000, ApJ, 531, L147

\bibitem{leinert01}
Leinert, C. {\it et al.} 2001, A\&A 369, 215

\bibitem{loinard03}
Loinard, L., Rodr\'{\i}guez, L. F. \&  Rodr\'{\i}guez,
M. I. 2003, ApJ, 587, L47

\bibitem{menard03}
M\'enard {\it et al.} 2003, ApJ, submitted

\bibitem{motte_andre01}
Motte, F. \& Andr\'e P. 2001, A\&A, 365, 440

\bibitem{myers98}
Myers, P. C. {\it et al.} 1998, ApJ, 492, 703

\bibitem{reipurth00}
Reipurth, B. 2000, AJ, 120, 3177

\bibitem{white_ghez01}
White, R. J. \& Ghez, A. M. 2001, ApJ, 556, 265

\bibitem{wood02}
Wood, K. {\it et al.} 2002, ApJ, 564, 887

\bibitem{zinnecker_wilking92}
Zinnecker, H. \& Wilking, B. A. 1992, in ``Binaries as Tracers
of Stellar Formation'', eds. Duquennoy \& Mayor, Cambridge
Univ. Press, p. 269

\end{chapthebibliography}
\end{document}